\documentclass{article}
\usepackage{spconf,amsmath,graphicx,url}

\title{Learning Filter Banks Using Deep Learning for Acoustic Signals}
\name{Shuhui Qu*, Juncheng Li*, Wei Dai, Samarjit Das}
\address{shuhuiq@stanford.edu, billy.li@us.bosch.com, wdai@cs.cmu.edu, samarjit.das@us.bosch.com}
\begin{document}
%
\maketitle
\begin{abstract}
Designing appropriate features for acoustic event recognition tasks is an active field of research. Expressive features should both improve the performance of the tasks and also be interpret-able. Currently, heuristically designed features based on the domain knowledge requires tremendous effort in hand-crafting, while features extracted through deep network are difficult for human to interpret. In this work, we explore the experience guided learning method for designing acoustic features. This is a novel hybrid approach combining both domain knowledge and purely data driven feature designing. Based on the procedure of log Mel-filter banks, we design a filter bank learning layer. We concatenate this layer with a convolutional neural network (CNN) model. After training the network, the weight of the filter bank learning layer is extracted to facilitate the design of acoustic features. We smooth the trained weight of the learning layer and re-initialize it in filter bank learning layer as audio feature extractor. For the environmental sound recognition task based on the \emph{Urbansound8K} dataset~\cite{salamon2014dataset}, the experience guided learning leads to a 2\% accuracy improvement compared with the fixed feature extractors (the log Mel-filter bank). The shape of the new filter banks are visualized and explained to prove the effectiveness of the feature design process.
\end{abstract}
\section{Introduction}
\label{sec:intro}

In the past few years, the research community has put significant efforts in designing feature representations for acoustic sound recognition. Good features should improve the performance of various audio analytic tasks such as classification and detection. Traditionally, features are heuristically designed, based on the understanding of spectral characteristics of natural sounds. Meanwhile, since this process is separate from the classification process~\cite{sainath2013learning}, heuristically designed features do not always contain enough information to obtain a high classification accuracy.

Thanks to the development of deep learning methods and rich dataset for sound, deep learning is increasingly becoming a popular candidate for acoustic recognition tasks~\cite{hinton2012deep,lecun2015deep,dahl2013improving}. Recently, CNN has shown the superior performance in feature extraction and classification in visual~\cite{he2015deep, simonyan2014very} and acoustic domain~\cite{sainath2015learning}, especially in speech recognition~\cite{simonyan2014very,sainath2015learning,he2015deep,lecun1995convolutional}. It could not only reduce the dimension of data and but also could extract  features as well. However, training a CNN model requires huge computational effort. Therefore, we leverage human experience (i.e. domain knowledge), to design the deep learning model, understand the  features from the model, and finally use the learned features to improve the audio recognition tasks' performance. 

\begin{figure}[htb]
\begin{minipage}[b]{1.0\linewidth}
  \centering
  \label{fig1:framework}
  \centerline{\includegraphics[width=8.5cm]{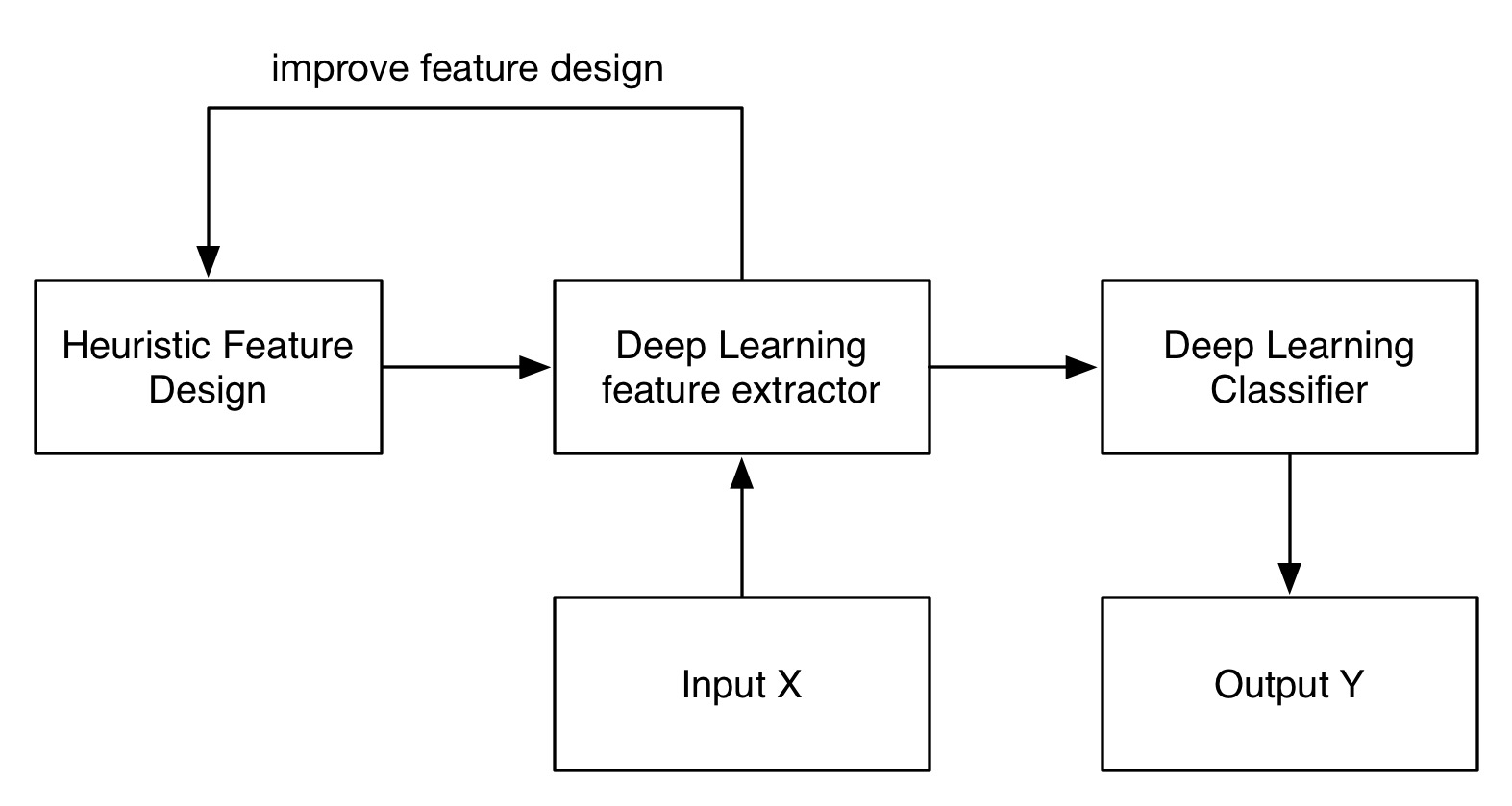}}
  \caption{Experience Guided Learning framework. The weight of the network is initialized by human heuristics. After training, the trained weight is post-processed and used to initialize the extractor. This process would iteratively improves the classification accuracy}\medskip
\end{minipage}
\end{figure}

Currently, most works in sound recognition area use log-mel filter banks as features. These features are not optimized for a particular audio recognition task at hand, and thus might not lead to high accuracy~\cite{sainath2013learning}. In this paper, we design our feature extractor by the studying the procedure of designing log-mel filter banks. We build a special filter bank learning layer and concatenate it with a CNN architecture. After training, the weight of the filter bank learning layer is post-processed with human experience. Then, the filter bank learning layer is re-initialized with the processed weight. The weight could be iteratively improved for feature extraction. This process is shown in Fig.~\ref{fig1:framework}. We call it as experience guided learning. To our knowledge, this is the first attempt to infuse domain knowledge of feature design to a deep learning pipeline for acoustic recognition tasks.

By using this method, the accuracy of recognition for the \emph{Urbansound8K} sound ~\cite{salamon2014dataset} increases at least 1.5\% accuracy based on the human designed filter bank under different settings, such as triangular window for MFCC.

The rest of the paper is structured as follows: Section 2 introduces the related work by using various methods to improve sound recognition tasks. Section 3 describes the special layer, a layer that could extract log-mel features and the CNN architecture in our work in detail. The experimental setup and result are shown in section 4. Finally, conclusion to our work can be found in Section 5.

\section{Related Work}
\label{sec:relatedwork}
There is a wide range of studies related with sound recognition, especially in speech recognition. \cite{rabiner1989tutorial} provided a detailed implementation of the Hidden Markov Model (HMM) on speech recognition by using the Linear predictive coding (LPC) features.~\cite{hirsch2000aurora} applied the HMM model on the MFCC features for speech recognition. With the advancement of deep learning, people applied different deep learning techniques, CNN in particular, for recognition.~\cite{lee2009unsupervised} applied convolutional deep belief networks to audio data and evaluated them on various audio classification tasks by using the MFCC feature. Their feature representations trained from unlabeled audio data showed very good performance. However, the MFCC feature is not generalized and not learned for improving different  task objectives.~\cite{sainath2013learning} thus proposed a filter learning layer to adaptively learn filter banks from the spectrum, and obtained good result in speech recognition. However, this learning layer is complex (multiple non-linear operations) and requires pre-estimation of the spectrum features' mean and standard deviation. Therefore, in this study, we propose a new filter learning layer based on the procedure of designing log-mel filter banks.

\section{Filter bank Learning Layer}
\label{sec:filterbank layer}
The mechanism of the filter bank learning layer is similar to the design of log-mel spectrogram~\cite{imai1983cepstral}, which has been widely used in automatic speech recognition. In general, there are several steps to calculate this feature:
\begin{enumerate}
\item Perform Fourier Transform to calculate power spectrogram
\item Apply the mel filter banks to all power spectrogram
\item Take the logarithm of all filter banks' energy
\end{enumerate}
Similar to this process, we design the network layer as following: the filter bank learning layer takes power spectrogram of a waveform as input. The layer generates the mel-features by multiplying the filters and individual spectrum. The number of filters is a hyper-parameter that represents the number of features to be learned.  After that, we take the logarithm of these features and input into a CNN architecture that has high performance in sound recognition. The filter bank learning layer's weight is not randomly initialized. Similar to triangular window or gamma-tone filter window, each row of the weight is activated once (non-zeros value) within a localized frequency range. 

Mathematically, the filter bank learning layer is described by the following equation:
\begin{equation}
m_{i,t} = W_i^Tf_{t} = \sum W_{i,j}f_{j,t}
\end{equation}
where $f_{t}$ is the individual power spectrum of the acoustic clip at time $t$, $W_i$ is the weight of $i^{th}$ filter bank. $j$ represents each individual element. This operation's output is the energy of the filter bank. 

Then, we take the logarithm of $m_{i,t}$ to get the log-mel filter coefficient for filter bank $i$
\begin{equation}
l_{i,t} = log(m_{i,t})
\end{equation}
Here, to prevent the $log$ non positive number error, the equation is further developed as:
\begin{equation}
l_{i,t} = log(Relu(m_{i,t}) + \epsilon)
\end{equation}
where Linear Rectified Units
$Relu(x)=\left\{
\begin{array}{c l}      
    x & x\geq0\\
    0 & x<0
\end{array}\right.$ and $\epsilon$ is a small constant (e.g. $1e-10$).
In order to optimize the objective function $L$, the filter bank learning layer's weight is gradually updated by taking the derivative of the objective function with respect to the weight. The update equation is:
\begin{equation}
W_{i} = W_{i} - \alpha \frac{\partial L }{\partial W_{i}}
\end{equation}
here, $\alpha$ is the learning rate and $L$ is the loss. By taking the derivative of the weight. The derivative function could be calculated through chain rule:
\begin{equation}
\begin{split}
\frac{\partial L }{\partial W_{i}}& = \frac{\partial L}{\partial l_i} \frac{\partial l_i}{\partial m_{i,t}} \frac{m_{i,t}}{W_{i}}\\
&= \frac{\partial L}{\partial l_i} \mathbf{1}{\{m_{i,t} > 0\}} f_{i,t}
\end{split}
\end{equation}
and here, $\frac{\partial L}{\partial l_i}$ is the loss gradient from previous layers.

The filter bank learning layer could adaptively extract features from the power spectrogram. Combining domain knowledge, the learned filter bank's weight could be further developed into generic filters. Different from \cite{sainath2013learning}'s work, our filter bank learning layer does not require estimating the mean and standard deviation of the input beforehand. Also, our method incurs less computation cost.

\section{Experiment}
\label{sec:experiment}
In this study, the training of the CNN model is performed on the natural sounds dataset , the \emph{Urbansound8k}~\cite{salamon2014dataset}. This dataset contains 8732 labeled sound excerpts ($<=4s$) of urban sounds from 10 classes: air conditioner, car horn, children playing, dog bark, drilling, engine idling, gun shot, jackhammer, siren, and street music. They are evenly divided into 10 folds. 

\begin{figure}[htb]
\begin{minipage}[b]{1.0\linewidth}
  \centering
  \label{fig2:arch}
  \centerline{\includegraphics[width=3.5cm]{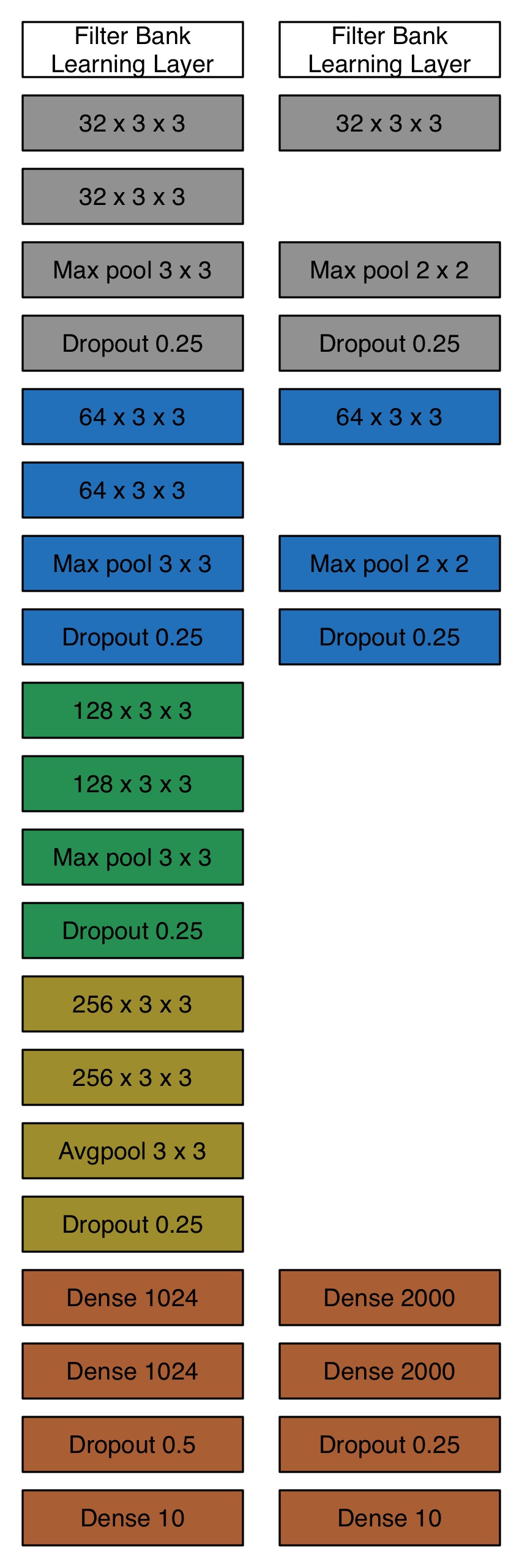}}
  \caption{We designed two architectures. The left model is a deep VGG CNN (arch 1), the right one is shallow architecture(arch 2)}\medskip
\end{minipage}
\end{figure}

The original sound is at 44.1kHz, we down sample it to 22.5kHz and 8kHz. For the 22.5kHz sound, due to the dimension of raw waveform, we divide it into 1 second each clip (in this case, we use majority voting method to obtain the output). After that, we take the power spectrogram of the sound by using libROSA~\cite{brian_mcfee_2015_32193}(nfft equals to sampling rate, default hop length). The weight of the filter bank learning layer is initialized by triangular filter banks of MFCC. We build two CNN architectures, one is deep VGG architecture~\cite{simonyan2014very} while the other one is shallow as shown in Fig.~\ref{fig2:arch}. The parameters are as following. The optimizer is default Adam optimizer~\cite{kingma2014adam} with learning rate 0.001. The learning rate decays every three epochs with the decay rate 0.006. The update function is:
\begin{equation}
lr = lr / (1 + decayrate \times epoch)
\end{equation}
where, $lr$ is the learning rate. 

After each layer, we apply leakyrelu\cite{DBLP:journals/corr/XuWCL15} with parameter 0.33.

The baseline is  around 70\%~\cite{salamon2014dataset} by using svm with rbf kernel and 73.7\% \cite{piczak2015environmental}. We also test the \cite{sainath2013learning}'s filter bank learning layer for comparison.

\section{Result and analysis}
\label{sec:resultAnalysis}
\subsection{Experiment Result}
\label{ssec:result}
The result is shown in Table 2. The proposed method could provide a modest 0.4\% of improvement in the classification accuracy. We take out the weight of the filter bank learning layer and use the Savitzky-Golay function~\cite{schafer2011savitzky} to smooth it. We then re-initialize the filter bank layer with the smoothed weight. After retraining the model, the accuracy is improved by 1.5\%. We didn't concanate the 4 second clip for the 22.5kHz sound, but we expect improvement compared to the 8kHz result. We also test the filter bank learning layer proposed in~\cite{sainath2013learning}, but the accuracy is lower than other baselines. This might be caused by the complex non-linearity of this layer and our estimation of input's mean and standard deviation might be too rough. To our knowledge, our method obtains the highest accuracy of \emph{Urbansound8K} dataset.

We also notice that the sampling rate of the sound affect the detection accuracy. For natural sound, different events happen at different frequency levels. Therefore, a relatively high sampling rate is essential for natural sound recognition tasks.

\begin{table}[t]
\caption{Result}
\label{resulttable}
\begin{center}
\begin{tabular}{lllllll}
\multicolumn{1}{c}{\bf Win} &
\multicolumn{1}{c}{\bf Arch} &
\multicolumn{1}{c}{\bf n\_filt} &
\multicolumn{1}{c}{\bf Weight} &
\multicolumn{1}{c}{\bf Dura} &
\multicolumn{1}{c}{\bf Freq} &
\multicolumn{1}{c}{\bf Acc}
\\ \hline \\
T &1 &128 &Fix  &1    &22.5      &71.88\\
T &1 &128 &Trained &1     &22.5      &72.21\\
T &1 &128 &Improved  &1       &22.5    &73.63\\
T &1 &128 &Fix  &4(MV)    &22.5     &78.34\\
T &2 &40 &Fix  &4    &8     &69.03\\
T &2 &40 &Trained  &4   &8     &69.43\\
T &2 &40 &Improved &4    &8     &71.41\\
\end{tabular}
\end{center}
\caption{T means initialized by triangular window; fix means fix the initial weight(in this case, the layer is the same as log-mel feature extractor), train means training with trainable weight, improve means smoothing the weight and reinitialize the filter bank learning layer with this weight. (MV) means post-processing using majority voting method.}
\end{table}

\subsection{Filter Bank Analysis}
\label{ssec:analysis}
The purpose of this work is to understand the mechanism of filter bank and further facilitate the design of filter banks to generate better feature extractors. Here, we visualize the filter banks from the triangular window, and smoothed weight from the trained filter bank learning layer that is trained by fold 1-9 in the following picture. 

As we can notice, the first few learned filter banks (1st row) conform with the triangle filter banks, which means these triangular filter banks capture most information in low frequency range. However, in the second row, we notice that the learned filter banks are activated around 0.4kHz to 0.5kHz and 0.75kHz to 0.9kHz, while frequency between 0.6 to 0.7 kHz is less interested. 
 
\begin{figure}[htb]

\begin{minipage}[b]{1.0\linewidth}
  \centering
  \label{fig1:filter}
  \centerline{\includegraphics[width=10cm]{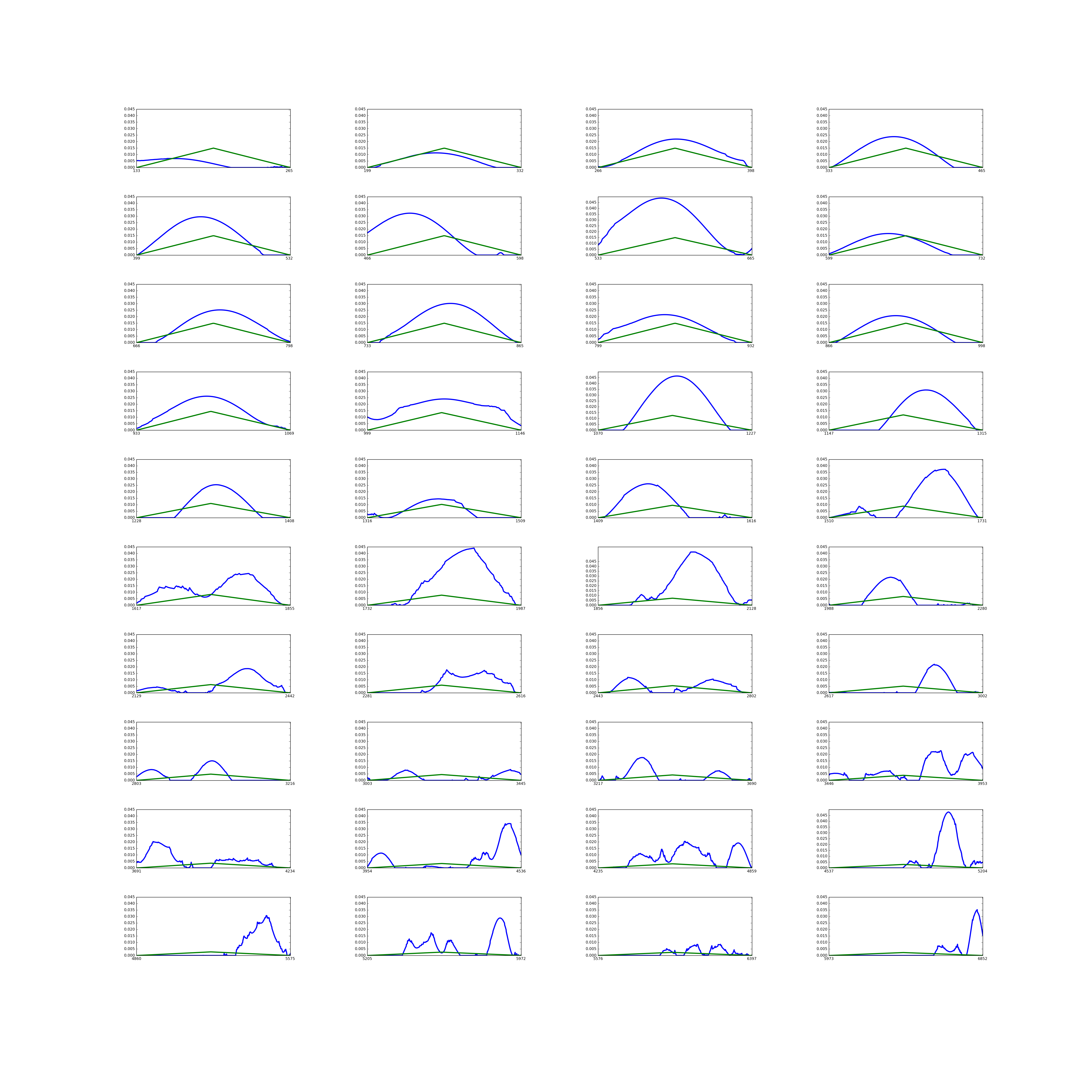}}
  \caption{Filter banks learned from Triangular Window, from top to down, from left to right is the increasing frequency band.}\medskip
\end{minipage}
\end{figure}

In triangular windows, the bandwidth of filters increases as the frequency level increases. Contrary to this, the learned filters show smaller bandwidth at relatively high frequency area. Fig.3 also shows there are several new peaks within the original single window, which means more filter banks are required. For instance, in the last row, the third picture shows that there are three different frequency ranges that are activated and their bandwidths are relatively small. This information could provide more intuition for audio experts to design new filters.

One problem with these learned filter banks is that they have a lot of serration along the the shape. This is primarily due to the bias of the model. By smoothing the learned filter banks, the model could be generalized, however, more expert experience would be beneficial to improve the recognition accuracy. Here, we apply the Savitzky-Golay function, however, different smooth function might result to different performance. Also, adding some regularization on model's parameters would smooth these filters as well.
\section{Conclusion}
\label{sec:conclusion}
In this paper, we explore the possibility of using the deep learning methods to facilitate the design of filter banks by incorporating human expert knowledge. We first design a filter bank learning layer that takes in frequency features. The output of the layer is fed to two different CNN architectures. This layer is designed according to the design procedure of the log-mel-spectrogram. By taking the weight of the filter bank learning layer, we apply a smooth function on the weight. This gives us at least 1.5\% accuracy improvement on the \emph{Urbansound8K} dataset. We further investigate the learned filter banks, and they provide us some intuitions to facilitate the feature design for the recognition task.

\bibliographystyle{IEEEbib}

\end{document}